\documentclass[runningheads]{llncs}
\usepackage{tikz}
\usepackage{graphicx}
\newcommand{\etal}{et al.$\!$}

\begin{document}
\title{Better Balance in Informatics:\\ An Honest Discussion with Students\thanks{Mariel's contribution was made while she was a Master's student at UiT.}}
%
%
\author{Elisavet Kozyri\inst{1} \and
Mariel Evelyn Markussen Ellingsen\inst{2}\and
Ragnhild Abel Grape\inst{1}\and
Letizia Jaccheri\inst{3}}
\authorrunning{E. Kozyri et al.}
%
\institute{UiT The Arctic University of Norway \and
Woid AS \and
Norwegian University of Science and Technology}
\maketitle              
\begin{abstract}
In recent years, there has been considerable effort to promote gender balance in the academic environment of
Computer Science (CS). However, there is still a gender gap at all CS academic levels: from  students, to PhD 
candidates, to faculty members. This general trend is followed by the Department of Computer Science at UiT The Arctic University of Norway. To combat this trend within the CS environment at UiT, we embarked on structured discussions
with students of our department. After analyzing the data collected from these discussions, we were
able to identify action items that could mitigate the existing gender gap at our department.
In particular, these discussions elucidated ways to achieve (i) a balanced flow of students into CS undergraduate program, 
(ii) a balanced CS study environment, and (iii)
a balanced flow of graduates into higher levels of the CS academia (e.g., PhD program).
This paper presents the results of the discussions and the subsequent recommendations that we made to the administration
of the department. We also provide a road-map that other institutions could follow to organize similar
events as part of their gender-balance action plan. 

\end{abstract}
\section{Introduction}

Innovations in Computer Science shape the lives of everyone in our society. 
To create innovative solutions tailored to everyone, it is important that all groups of society are represented in the creation of these solutions.
However, this is still not the case in the field of Computer Science (CS).
Having an awareness of the lack of representation and the different barriers people face in CS are
fundamental in helping the field target those challenges and becoming more equitable and inclusive~\cite{TDT10-14}.

Statistics from
Europe show that women are still highly underrepresented
in CS. According to Eurostat~\cite{eurostat}, the percentage of female specialists in 
Information and Communications Technology has evolved from 17\% in 2012 to 19,1\% in 2021. At university level in STEM, the percentage of female Bachelor, Master, and PhD students is 20\%, while the percentage of female professors is 15\%.

Specifically for the Department of Computer Science at UiT The Arctic University of Norway,
only $13\%$ of students, $14\%$ of PhD candidates and  $21\%$ of faculty members are female.

\emph{Better Balance in Informatics} (BBI), a program led by the CS department at UiT and funded by the Research Council of Norway, aims to rectify this imbalance and create a more diverse learning environment for Computer Science.
BBI is connected and builds upon an ecosystem of national and international projects which address gender balance in CS acting on different levels: school (\cite{happe21}, \cite{eugain-booklet}),  university (\cite{prestige}, \cite{balansehub} \cite{Jaccheri2022}), industry (\cite{diversityEU}, \cite{google}, \cite{microsoft}), and the interplay of these levels (\cite{eugain}). 

BBI aimed to identify some of the reasons that led to the current gender dynamics in our CS department, and then propose measurements that could address those reasons.
Hearing directly from the CS students (Bachelor, Master) seemed to be a sensible way for us to identify those reasons.
So, BBI organized structured discussion sessions, where we invited CS students (Bachelor, Master) to share their thoughts about: 
\begin{enumerate}
\item the reasons they picked CS for their studies, 
\item their current experience with the CS studies, 
\item their intention to pursue an academic career in CS, and 
\item ways to make the CS community more diverse and inclusive.
\end{enumerate}
The answers of the students illuminated points of intervention, which could lead to
a balanced flow of students into CS undergraduate program, 
a study environment that embraces diversity, and 
a balanced flow of students into higher levels of the CS academia.
 
This paper presents the methodology (\S \ref{sec:methodology}) we employed to organize the discussion sessions, to collect responses, and to report the results. We then present the specific questions we asked the students and the analysis of their answers (\S \ref{sec:results}). Finally, we list the recommendations (\S \ref{sec:recommendations}) submitted to the CS department for achieving a gender-balanced environment,
we discuss related work (\S \ref{sec:relatedwork}), and we conclude (\S \ref{sec:conclusion}) with reflections about the discussion sessions.

\section{Methodology}\label{sec:methodology}

The end goal of the discussion sessions was to identify points of interventions that could increase the gender balance among the incoming CS students, the current CS students, and the CS graduates that are interested in entering the CS academia. 
To identify those points, we were aiming for a high number of
participants in the discussion sessions: the more participants, the greater the plurality of experiences, and thus, the higher 
the chances to find opportunities for improvement.
Deciding which questions to ask was crucial to ensure that experiences from different aspects of the CS studies are captured and then analyzed.
But, we also had to create a trusting discussion environment 
for the students to honestly share those experiences with us.
This section describes the
methodology we followed to prepare and organize the discussion sessions such that all these targets are met.

\subsection{Outreach}

Attracting the attention of students and persuading them to participate in the discussion sessions was not trivial.
Unless there is an immediate academic or employment gain, motivating students to devote part of their busy
schedules to a university-led event is indeed challenging.
Our strategy to address this challenge comprised the following steps:

\paragraph*{Hiring students as project assistants.}
We hired two talented and enthusiastic female students as assistants for the project.
They were our bridge to the student community in our CS department.
And this bridge was functioning in both ways. Their thoughts, insights, and experience informed all aspects
of the BBI project, including the questions we asked during the discussions.
At the same time, they knew best how to reach their fellow-students and promote the agenda of BBI (e.g., what advertisement means to employ and what to say in these means).

\paragraph*{Website.}
The BBI website (\url{https://uit.no/project/bbi}) is the main official space where the mission of BBI is communicated
to the world.
So, we redesigned this website to include a clear and short motivation for the BBI mission, and describe
the upcoming BBI events, in particular the discussion sessions.

\paragraph*{Advertisement.}
To reach a wider set of students and persuade them to participate in the BBI discussion sessions, we employed a variety of means. 
We posted advertisements on the monitors of the Department, the social networks of the Department, on Canvas, the UiT calendar, and the local student organization forum, which is a Discord server that is maintained by the student organization TD.
The student assistants also gave 5-minutes talk about BBI and the discussion
sessions to courses with high enrollment, 
they created and distributed flyers, and
they organized a stand with coffee and cookies, where
students could casually socialize and talk about BBI. In terms of registrations to the BBI discussion sessions,
Canvas and TD seemed to have been the most effective, since we observed a high correlation between the time a post about
the BBI event was created and the time students were registered.

\paragraph*{Open to everyone.}
The invitation to participate in the BBI discussion sessions was open to all students of the CS department,
independently of their gender (female, male, non-binary).
This is because the gender imbalance is a problem that owes to concern everyone---not only a part of the community.
And because any further actions that the Department will take to address the problem might effect every student,
there needs to be a wider understanding that these actions are worthwhile.
Leaving specific groups of students outside the discussion, 
would not have increased this understanding.

\subsection{Discussion Sessions}

The discussion sessions were held at Árdna, UiT. 
Árdna is an iconic place in the university, ideal for secluded discussions.
Its central fire place and the surrounding wooden benches invites people to open up and discuss honestly.

In the BBI discussion sessions participated around 20 participants.\footnote{We do not give specific numbers to preserve
the anonymity of the participants.}
Comparing to events organized in the past by BBI, this number of participants was a very welcoming surprise. 
From those participants, around $50\%$ were female or non-binary students, and around  $50\%$
were male students. The vast majority of the students participated physically, but there were some that participated remotely.
There were around three participants per discussion session.
Each discussion session was moderated by two BBI members: one member was asking the questions, and the other member
was typing the answers (we used no video or sound recording).
At least one of the moderators was always one of the BBI student
assistants; having participants talking to their peers led to frank discussions.
To preserve anonymity, each participant was assigned a number, so the recorded answers were associated with these numbers---not with the identity of the student.
For the discussion sessions, we gave the option to the student to select either Norwegian or English as the speaking language.
All participants, apart from those that participated remotely, were offered a full meal and a free cinema
ticket.

\subsection{Selection of Questions}

The selection of questions was inspired by questionnaires developed by other gender-balance projects, such as EUGAIN~\cite{eugain},
Prestige in UiT~\cite{prestige}, and BalanseHub~\cite{balansehub}. However, the questions were tailored for the needs of the CS department in UiT. And, in particular, the questions were intended to
cover a wide range of students' experience: from the point they considered applying for CS, to their current studies and their future plans.

\subsection{Reporting of Results}

For most of the questions asked during the BBI discussion sessions, we compiled the answers into graphs.
Each graph depicts how answers are correlated with different genders. This information help us guide
our selection of action items for improving the gender balance.
We protect the anonymity of the participants, so we do not give specific numbers at graphs.
Also, we do not have a separate category for non-binary participants, because the number of non-binary participants was not high enough to protect their anonymity.
Instead, we group female and non-binary participants together, and we explore the dynamics between majority (males) and minorities (female, non-binary).

\section{Results}\label{sec:results}

This section presents the questions we asked the participants and their answers concerning: 
\begin{enumerate}
\item the reasons they picked CS for their studies, 
\item their current experience with the CS studies, 
\item their intention to pursue an academic career in CS, and 
\item ways to make the CS community more diverse and inclusive.
\end{enumerate}
Correlating their answers with their gender, we identified action items that could lead to
a balanced flow of students into CS undergraduate program, 
a study environment that embraces diversity, and 
a balanced flow of students into higher levels of the CS academia.

\subsection{Intention to Study CS}

\begin{figure}[t]
\centering
\includegraphics[width=\textwidth]{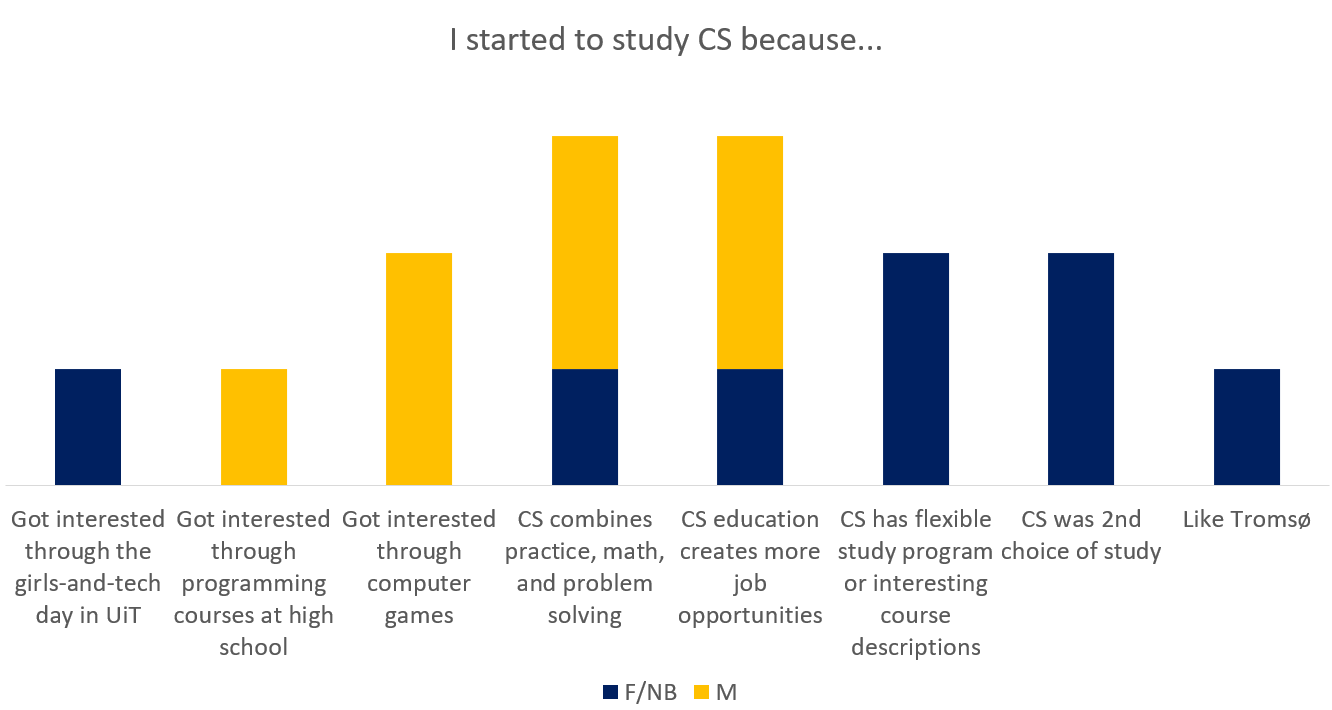}
\caption{Reasons for choosing to study CS. Each column corresponds to a different reason. The height of a column represents the number of participants that submitted the corresponding reason. Dark blue represents female or non-binary participants (F/NB); yellow represents male participants (M).}
\label{fig1}
\end{figure}

To increase the balance in Computer Science, one first needs to increase the balance in the new-coming
students. So, when advertising CS to younger students, one could also include aspects that attract
minorities. We tried to identify those aspects by asking the participants the reason they decided
to study CS in the first place.
Figure \ref{fig1} shows the different answers we received. The higher the column, the more students
gave that answer. The graph also shows how the answers are split between the minority (F/NB) and majority (M).
There is a correlation between the gender and the reason for selecting CS studies.

\paragraph*{\textbf{Action Items.}} Observing Figure \ref{fig1}, we can
identify the reasons the minority chose CS studies:
the problem solving aspect of CS, the flexibility
of the CS studies, the job opportunities that CS graduates enjoy. To increase the diversity of incoming students, we
can then emphasize those reasons when advertising CS.
Also, as a possible means of advertisement Figure \ref{fig1} indicates the UiT girls-and-tech day.

\begin{figure}[h!]
\centering
\includegraphics[scale=0.65]{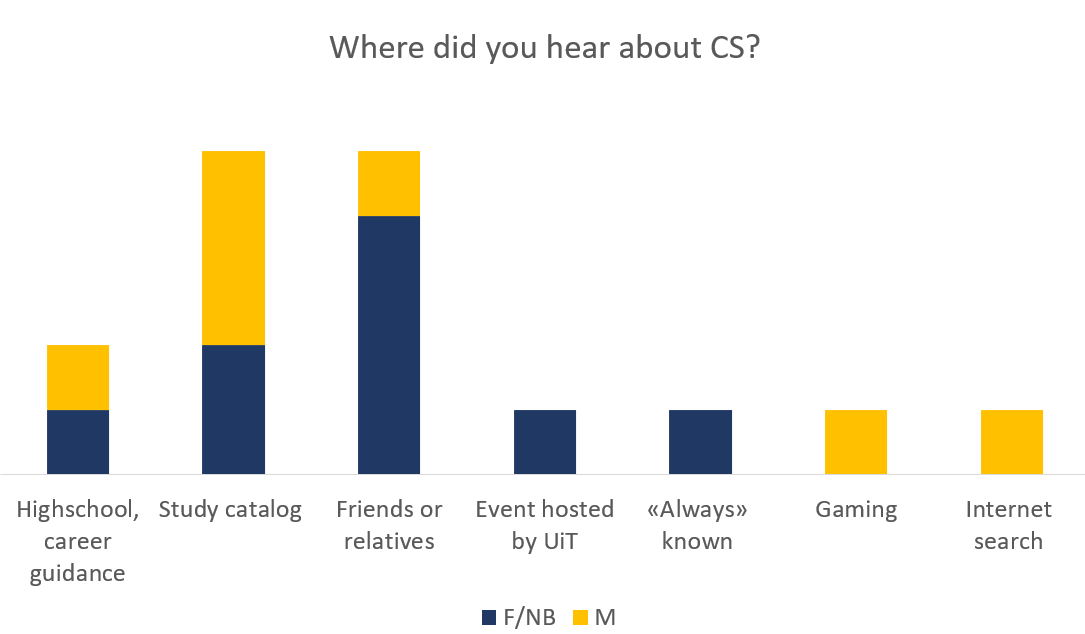}
\caption{Ways for becoming familiar with CS.}
\label{fig2}
\end{figure}

Apart from the UiT girls-and-tech day, we wanted to understand what would be other effective means of
advertisement for attracting minorities to CS studies. So, we asked the participants where did they hear about CS.
Figure \ref{fig2} plots the answers, which are again correlated with the gender.

\paragraph*{\textbf{Action Items.}} Figure \ref{fig2} indicates that one could use the highschool and the university's study
catalog to better promote CS to minorities. Interestingly, friends and relatives have a high impact on the decision
of minorities to study CS. So, one can make tech employees ambassadors of CS to young female and non-binary members of
their families.

\begin{figure}[h]
\centering
\includegraphics[scale=0.3]{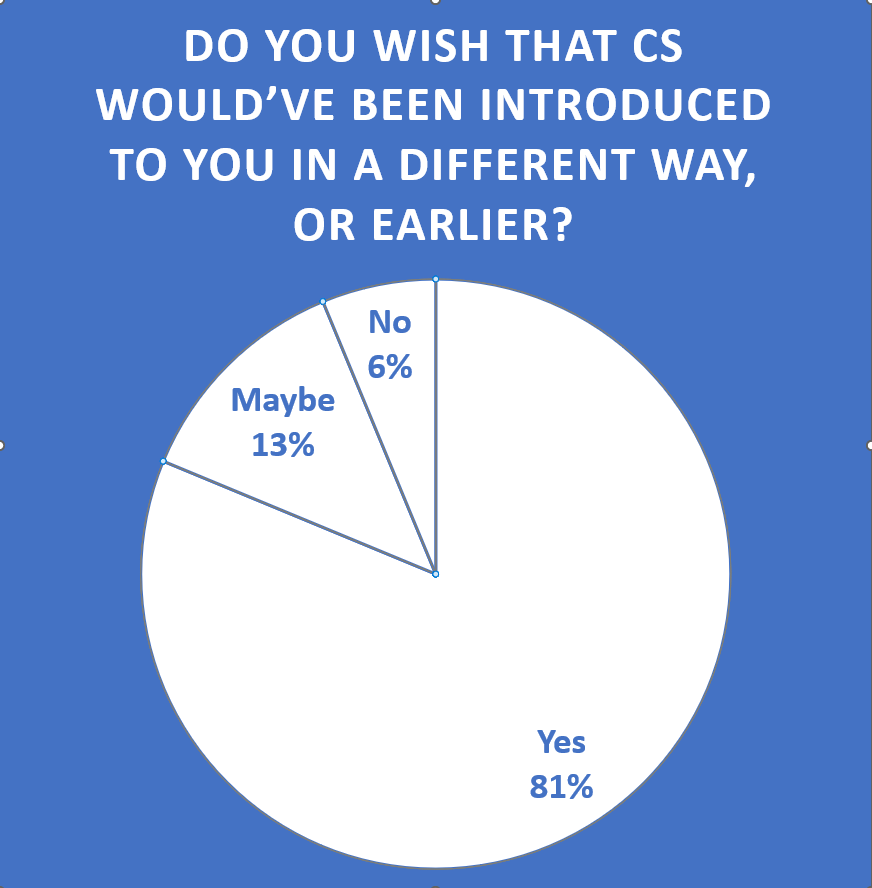}
\caption{Independent of their gender, $81\%$ of the
participants said that they would have liked to be introduced to CS in a different way.}
\label{fig3}
\end{figure}

In general, the vast majority of the participants, independently of their gender,
would have liked CS to have been introduced differently to them, as
Figure \ref{fig3} indicates. Participants indicated that CS should be introduced as something
that everyone can do and something that offers a plausible path to a regular job.

\paragraph*{\textbf{Action Item.}} When advertising to high-school students, we need to break stereotypes on who can study CS.

\subsection{Your Experience in the Current CS Environment}

\begin{figure}[h]
\centering
\includegraphics[width=\textwidth]{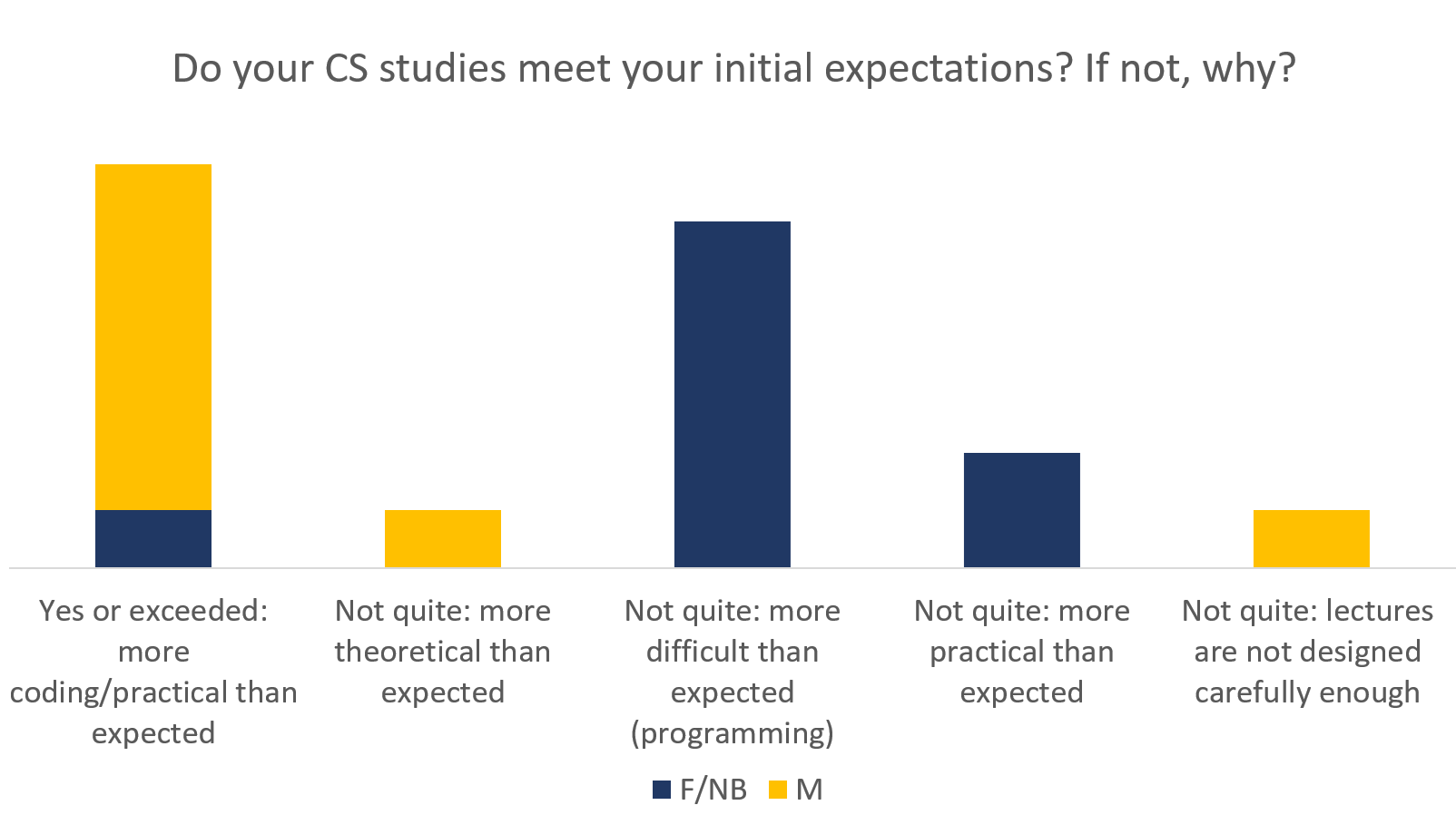}
\caption{The first column corresponds to the answer that the CS studies met or exceeded the expectations of the participant. The remaining four columns correspond  to the answer that the CS studies did not quite meet the expectations of the participant, and they also represent
different reasons why. 
The height of a column represents the number of participants that submitted the corresponding answer. Dark blue represents female or non-binary participants (F/NB); yellow represents male participants (M).}
\label{fig9}
\end{figure}

At the very least, we want to sustain the diversity among the incoming students, while these students
progress to more senior years; we aim to decrease the number of
drop-outs, with an emphasis on the minority group.
To achieve this, we need to assess the student's experience within the CS department
and identify aspects that can be improved to accommodate gender diversity.

We start by asking the participants whether the CS studies met their initial expectations.
Their answers are depicted in Figure \ref{fig9}.
Almost all minority participants gave a negative answer: they found their studies either
more difficult or more practical than expected. In particular, they found the learning curve
of programming to be particularly steep, something that might drive some of the minority students to drop-out.
We believe addressing this concern is important for maintaining a diverse environment in the department.

\paragraph*{\textbf{Action Item.}} We propose the adoption of a smoother introduction to programming,
which will be appropriate for students with no  prior programming experience. 

\bigskip
\noindent
Returning to Figure \ref{fig9}, one also notices that, for most of the male participants,
their experience in the CS studies met or even exceeded their initial expectations.
So, this question highlights a striking difference between the minorities and the male students
in terms of how they view their studies. This difference might be a cause of the 
current gender imbalance in the department.

All the participants agreed, though, that the social environment built around their studies
exceeded their initial expectations. This is a great achievement of the department that needs to be preserved
for the future, too.

\begin{figure}[h!]
\centering
\includegraphics[scale=0.3]{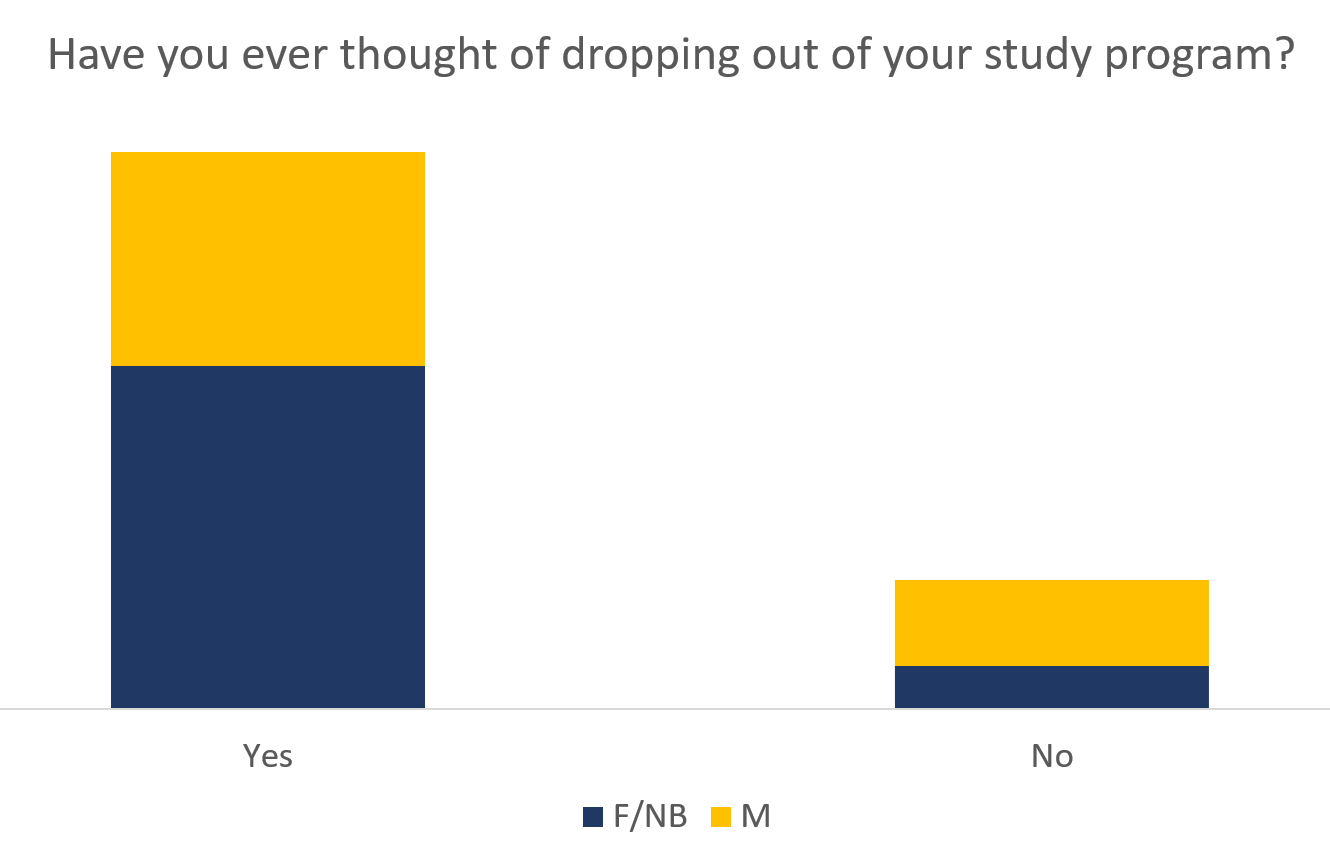}
\caption{The majority of the participants replied that 
they have thought of dropping out of their CS study program.}
\label{fig10}
\end{figure}

Participants were then explicitly asked whether they have thought to drop-out of their study program.
As Figure \ref{fig10} shows, most of the participants answered affirmatively.
Notice that this is a concern across all genders, opposing the misconception that the minorities
are more likely to have such thoughts. Notice also that even though most of the male students thought
to drop-out of the program, they still had an overall positive assessment of their study experience,
according to Figure \ref{fig9}. 

\begin{figure}[h!]
\centering
\includegraphics[scale=0.3]{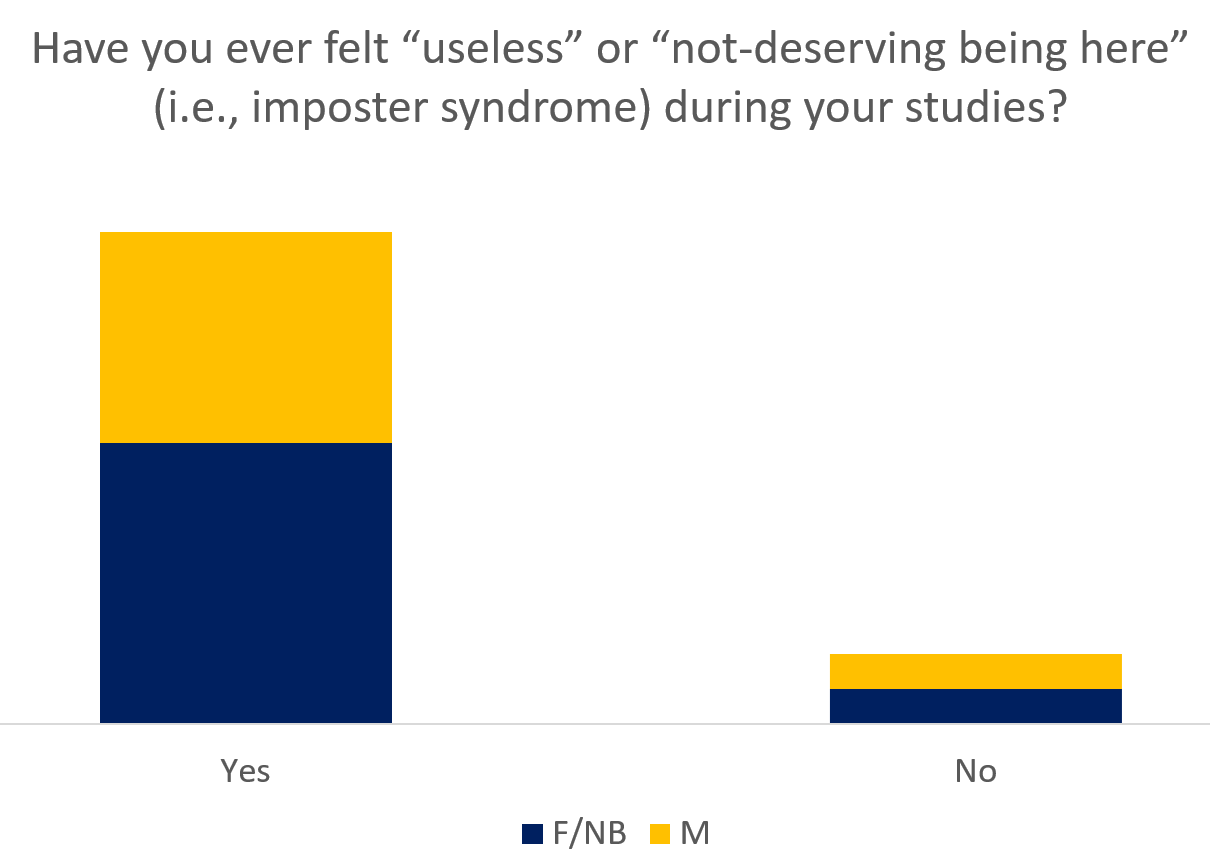}
\caption{The majority of the participants replied that 
they have have felt ``useless" or ``not-deserving to be here" during their CS study program.}
\label{fig11}
\end{figure}

As reasons for thinking to drop-out, the participants cited the difficulty of some courses,
the time-consuming assignments with overlapping deadlines, the demanding task of writing thesis (a task for 
which they did not feel prepared), and the complications that the COVID-19 pandemic brought.
For a student to be thinking to drop-out, it means that the student's self esteem might be low at that
point. Figure \ref{fig11} validates this claim, showing that
most of the participants felt "useless" or "not-deserving" being in the CS program.
Again, the answers do not seem to be correlated with the gender.
However there is an underlying difference: many of the males had this feeling once,
related to a specific assignment or for short period of time, whereas
the minority students had this feeling for a long period of time (i.e, months or even years). 

When asked about the reasons they instead decided to stay in the program and not drop out, the participants mentioned:
\begin{itemize}
\item the robust social network that they have built within and outside UiT, where they could talk about
their struggles,
\item the senior student advisor Jan Fuglesteg,
\item their self-determination and discipline,
\item taking time to relax.
\end{itemize}

\paragraph*{\textbf{Action Items.}}
Given the stimulating power that the social groups exercised on the students, we should further support
actions and groups that promote social networking in the department. Also, we will organize events
where senior students can offer tips and tricks from their experiences to the junior students, where the main
message will be ``if we have made it, so can you".

\begin{figure}[h]
\centering
\includegraphics[scale=0.3]{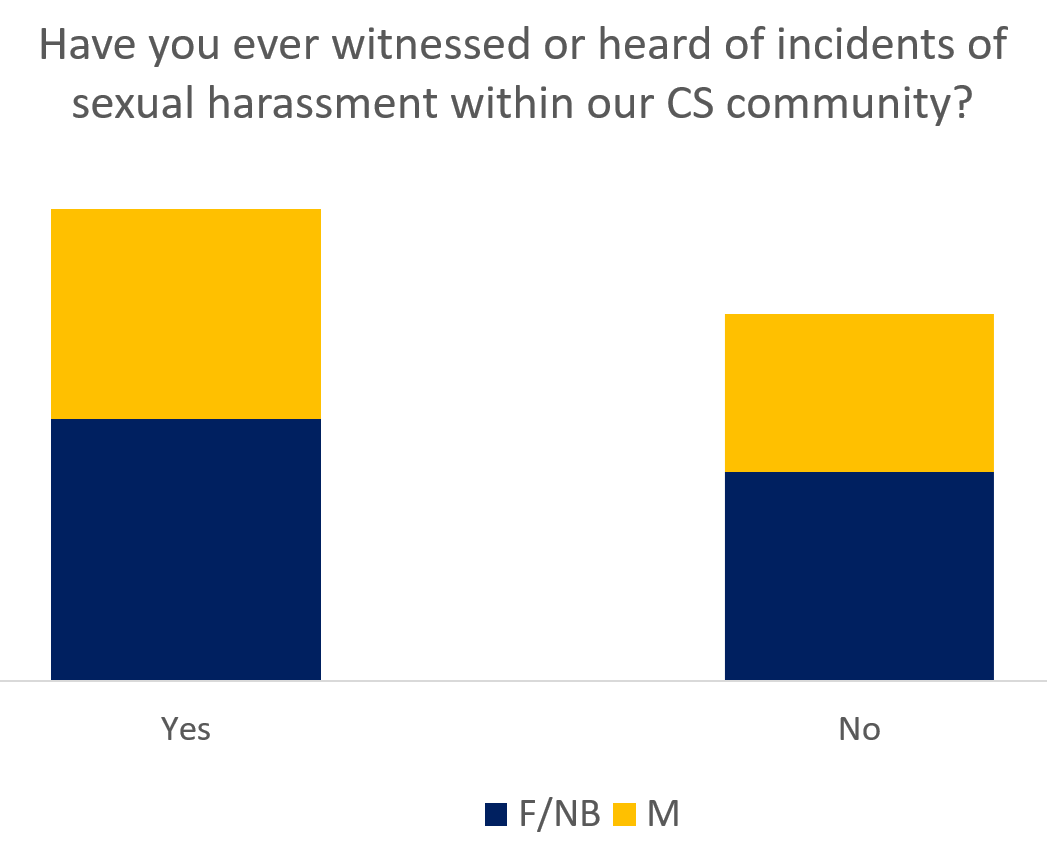}
\caption{More than half of the participants said that they have witnessed or heard of sexual harassment incidents
within our CS community.}
\label{fig12}
\end{figure}

\bigskip
\noindent

Concentrating on minority students, one of the reasons they might feel uncomfortable in an environment (and ultimately drop-out of the program) is when they have experienced sexual harassment.
So, we asked the participants whether they have ever witnessed or heard of sexual harassment incidents within the CS community. Figure \ref{fig12} depicts their answers.
More than half of the participants answered positively. 

\paragraph*{\textbf{Action Item.}} The ``Yes" column in Figure  \ref{fig12} should not exist. So, we will propose to the department to devise videos and examples of unacceptable behavior, so the student can recognize and dissociate from these phenomena.

\bigskip
\noindent

The experience that a student gets from their CS program is a collection of many educational aspects, such as
lectures, colloquiums, and assignments. For the CS program to be inclusive and diverse, all these aspects
should promote inclusiveness and diversity. We asked the participants if they feel that the
educational aspects below promote only a particular way of thinking.
\begin{itemize}
\item Lectures: Participants mentioned that examples employed in some lectures are more appealing to male students. These examples usually involve games or cars.
\item Colloquiums:\footnote{A colloquium is associated with a course and it is often led by a TA, who answers student's questions about the course material and assignments.} Participants, from all genders, mentioned that a colloquium can quickly get a
``boys-club" atmosphere if the TA is not attentive enough. The participants also express the wish for more
female TAs. 
\item Assignments: Some assignment are very focused on gaming or promote competitiveness, which might be
uncomfortable for some students. 
\end{itemize}

\paragraph*{\textbf{Action Items.}}
We will advise the faculty members to ensure that lectures and assignments accommodate a variety of interests.
The department should organize a seminar for TAs in which they become sensitized on not allowing colloquiums to 
become ``boys-clubs" and accommodating different interests and needs. We also need to brainstorm on how to hire more female TAs.

\subsection{Intention to Enter Higher Levels in CS Academia}

\begin{figure}[t]
\centering
\includegraphics[scale=0.3]{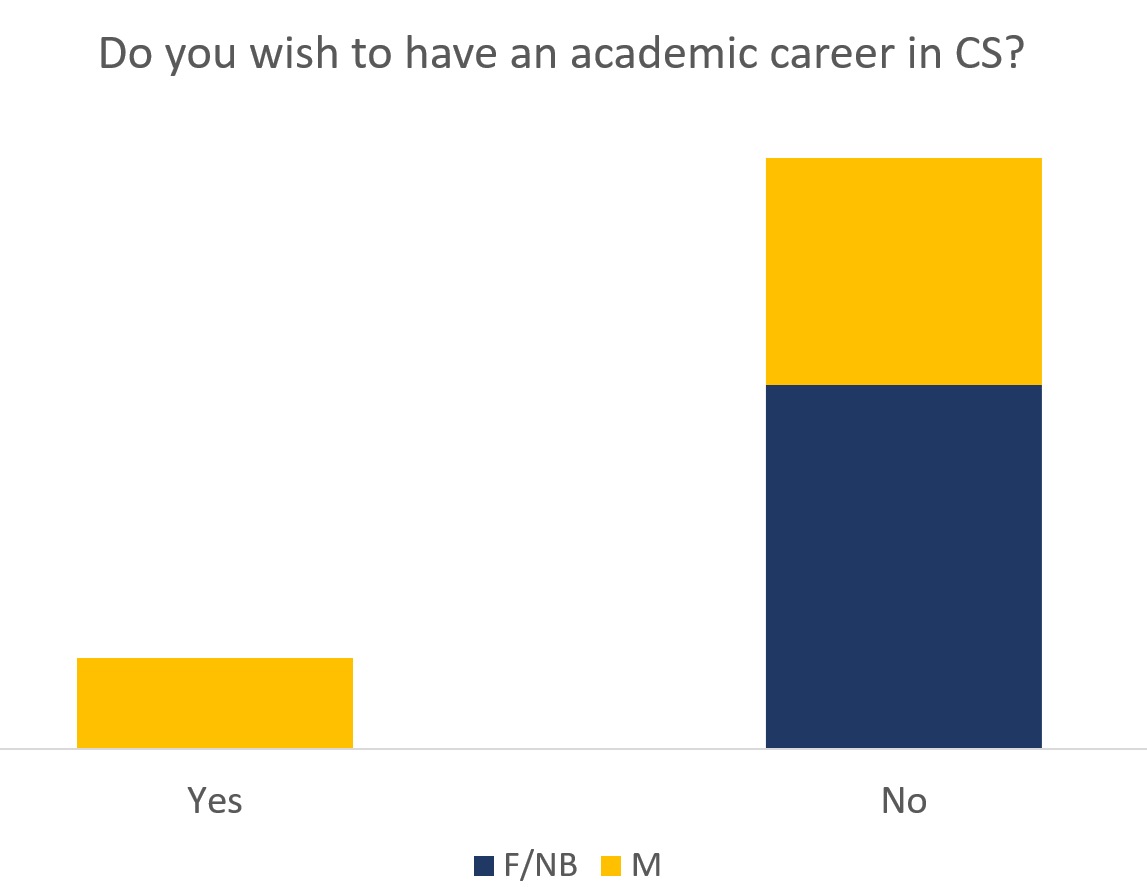}
\caption{Almost all of the participants do not wish to have an academic career in CS.}
\label{fig4}
\end{figure}

We are striving to achieve balance at all levels of the CS academia, from students to professors.
At this section, we focus our attention to higher levels in CS academia (from PhD candidates and above),
and we want to understand the intention of the current students to enter that level.
According to Figure \ref{fig4}, the vast majority, and in particular all female and non-binary participants, do not
wish to have an academic career in CS. Here are some reasons why:
\begin{itemize}
\item The academic career seems difficult or exhausting. 
\item No interest in research or writing or teaching.
\item Preference for work-life balance offered by the industry (pay, social, use tools, practical).
\item The description of an academic career is not clearly communicated.
\end{itemize}

\begin{figure}[h!]
\centering
\includegraphics[scale=0.3]{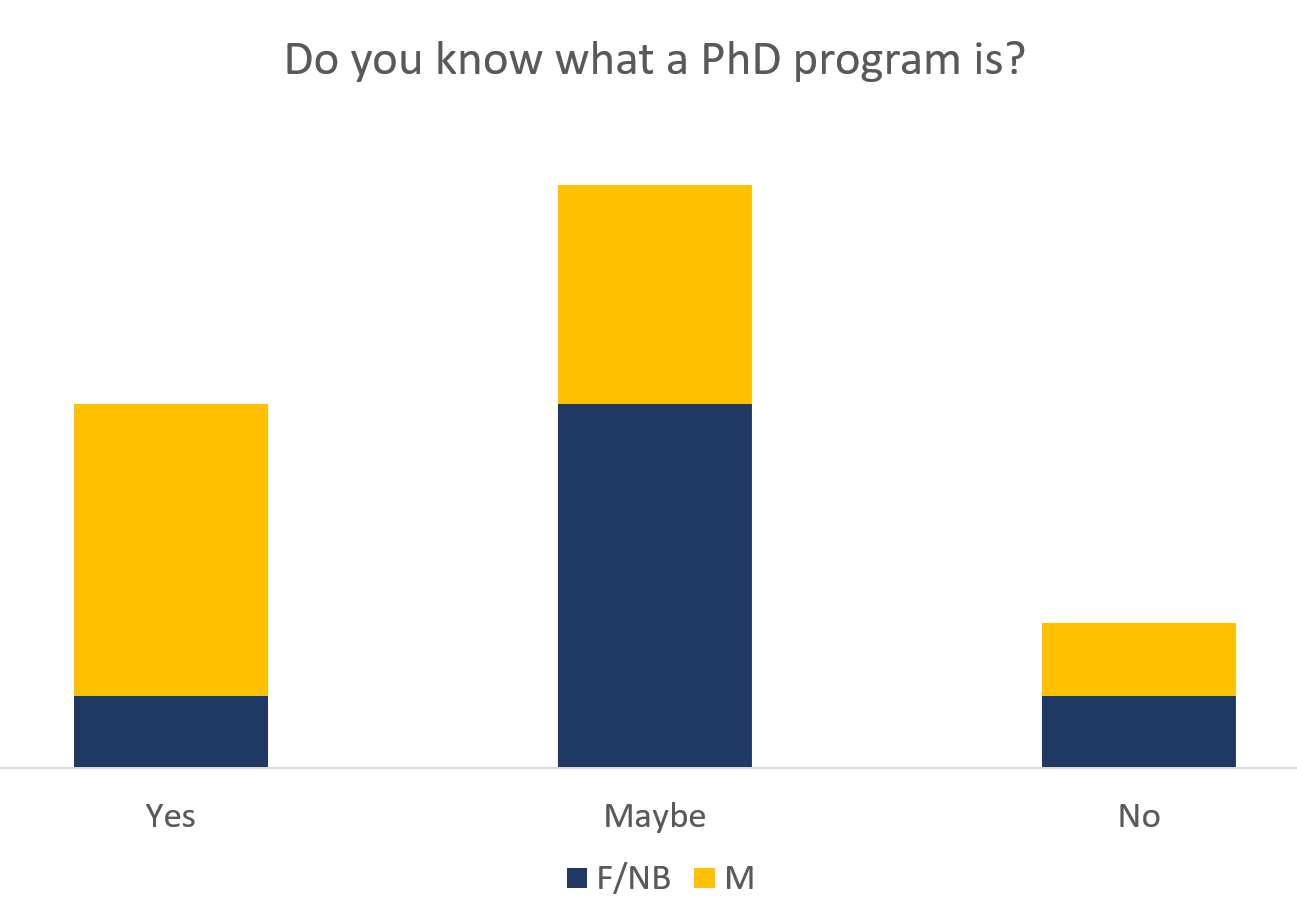}
\caption{Most of the participants are either unsure about or do not know what the CS PhD program.}
\label{fig5}
\end{figure}

\begin{figure}[h]
\centering
\includegraphics[scale=0.3]{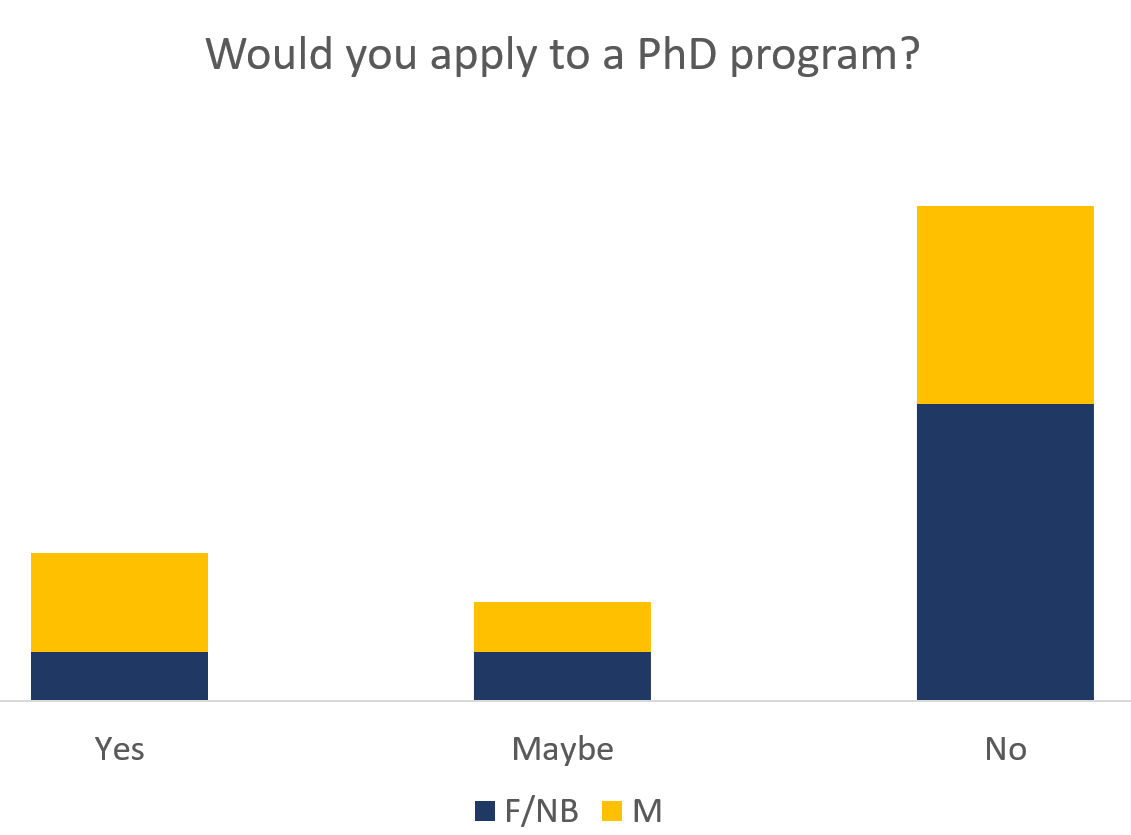}
\caption{Most of the participants would not apply to a PhD program.}
\label{fig6}
\end{figure}

In fact, as indicated by Figure \ref{fig5}, there is uncertainty between the students, and in particular
the minority students (i.e., female and non-binary), about what a PhD program is---the first stepping stone towards building an
academic career.
Given this uncertainty, it is expected that many students will not apply to a PhD program, something
that is affirmed by Figure \ref{fig6}.
Notice, though, that comparing Figures \ref{fig4} and \ref{fig6}, the participants are less negative towards
pursuing a PhD program than ultimately following an academic career.
This is because, after obtaining a PhD degree, there is always the possibility to follow a career in the 
industry. And some of the participants that replied ``no" to the prospective of applying to
a PhD program now, they contemplate the possibility of applying after working in the industry.

\begin{figure}[h!]
\centering
\includegraphics[width=\textwidth]{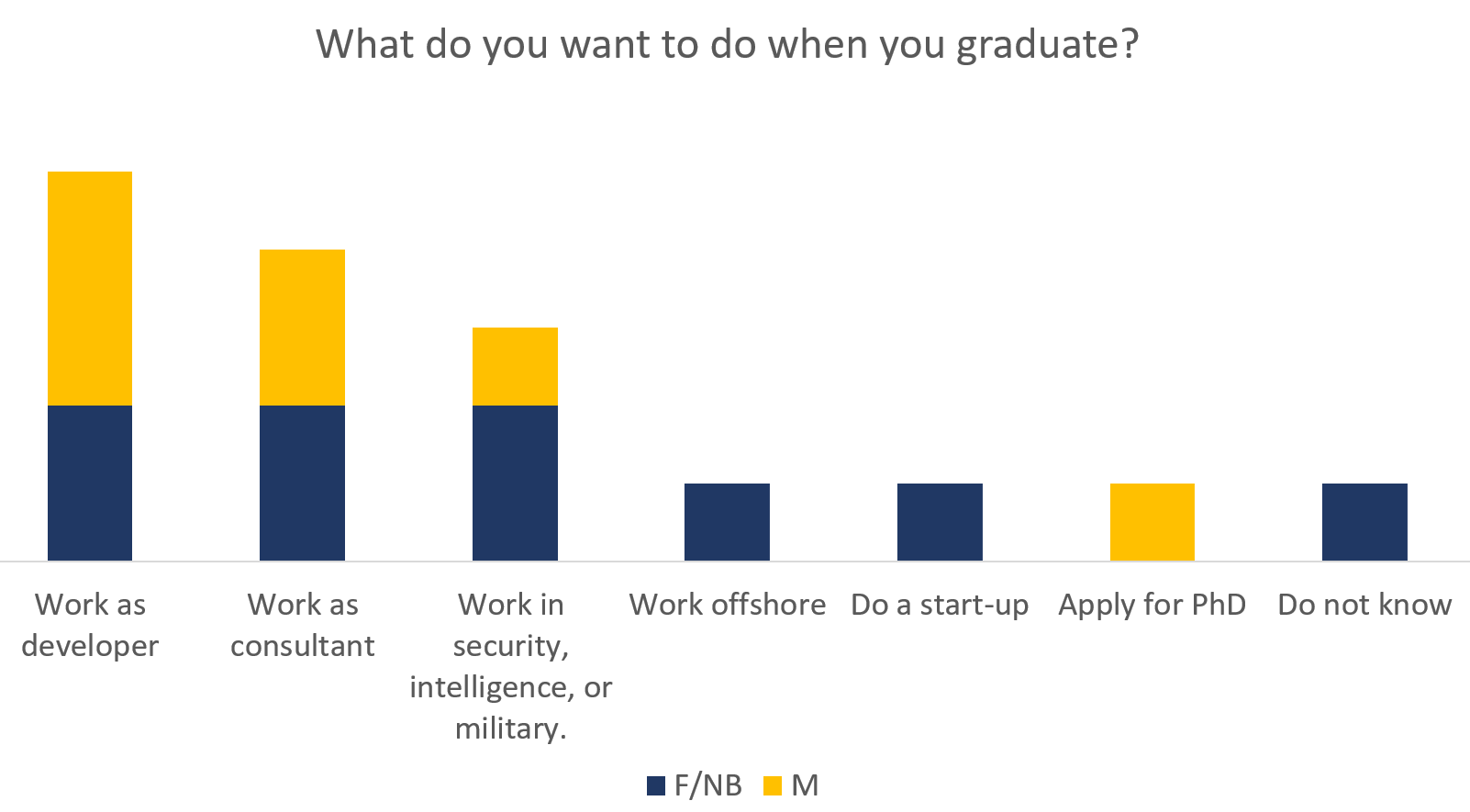}
\caption{Columns correspond to different options the 
participants consider to follow after graduation.}
\label{fig7}
\end{figure}

And if a participant does not want to apply to a PhD program, what are their immediate plans after graduation?
Figure \ref{fig7} answers this question. Notice that participants from the minority group
explore a wider variety of options.

\begin{figure}[h]
\centering
\includegraphics[width=\textwidth]{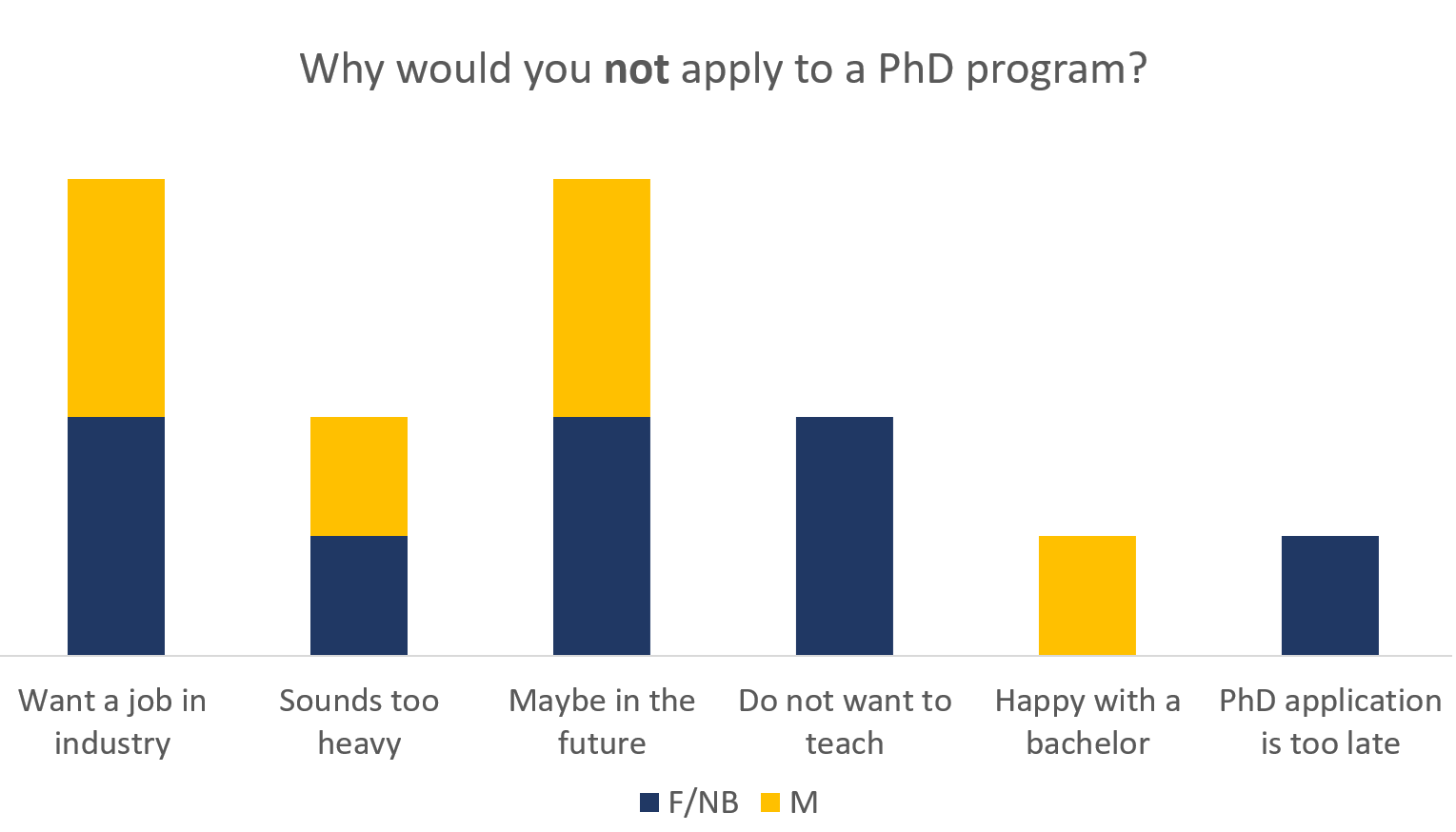}
\caption{Columns correspond to different reasons why the 
participants are not considering to apply to a PhD program.}
\label{fig8}
\end{figure}

Also, for the participants that are not considering to apply to the CS PhD program, Figure \ref{fig8} gives
the main reasons behind this disposition. Figure \ref{fig8} informs how we can intervene and address 
some of these reasons, possibly leading to more students applying to our PhD program. 

\paragraph*{\textbf{Action Items.}}
According to Figure \ref{fig8}, some participants said that the PhD program ``sounds too heavy", and they described PhD students as being ``alone" and ``depressed". While this description might portray one aspect of the PhD experience, it is definitely not the entire truth. So, we are going to hold events that clearly describe the characteristics of a PhD program,
emphasizing the positive aspects of being a PhD student. These events will also address the uncertainty that
was surfaced in Figure \ref{fig5} about what is a PhD program.

The late deadline for applying to a PhD, which is not synchronized with the job-search period of senior students,
is another reason why current students do not select a PhD program.
To remedy this, we will advise the faculty members of the CS department to consider moving the 
PhD application earlier in in the academic year (i.e., fall semester).

Finally, given that many participants said that they might consider applying to a PhD program in the future (i.e., after acquiring some experience in the industry), we advocate to advertise new PhD positions explicitly to
CS alumni. For some of these alumni, these positions might seem attractive.

\subsection{The Gender Gap and Possible Measurements to Close it.}

\begin{figure}[h]
\centering
\includegraphics[width=\textwidth]{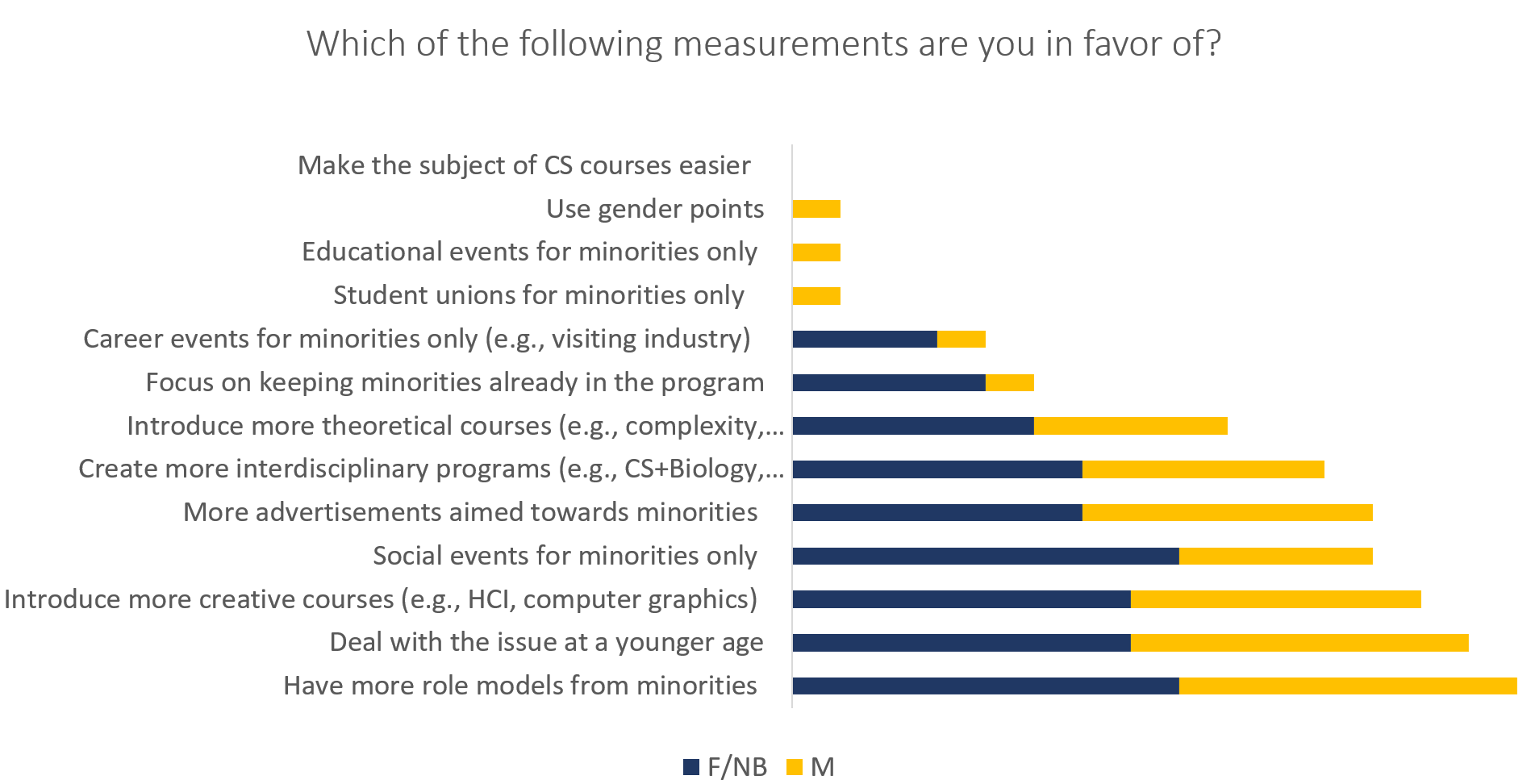}
\caption{Each row corresponds to a different measurement
for improving gender balance in CS. 
The length of each row represents the number of participants that agree with the corresponding measurement. Dark blue represents female or non-binary participants (F/NB); yellow represents male participants (M).}
\label{fig13}
\end{figure}

In previous sections, we attempted to understand the reasons why a gender imbalance exists in the CS department, and concluded with action items that could address those reasons.
In this section, we explicitly discuss gender balance with the students and record their opinions and proposals
on the subject.

To begin with, the vast majority of the participants said that the current gender-imbalance in the department
is a problem. They actually see that gender-balance has advantages: promotes innovation, enables plurality of
perspectives, and leads to a better study and work environment.
Many of the participants said that there are no disadvantages with gender balance, although some expressed
the concern that gender-balance efforts, such as gender quotas, might ``lead to people being hired for the wrong reasons".

The participants were then presented with different measurements that could be employed to improve the gender
balance within the department and asked to say whether they are in favor or not of each presented measurement.
Figure \ref{fig13} depicts their answers.
Notice that measurements that blatantly favor minorities in the education or in the career were among the least popular (for both minority and male participants).
We aim to focus on the most popular measurements.

Participants also proposed two additional measurements that did not appear in our pre-selected list:
\begin{itemize}
\item Share success stories from minorities.
\item Have a few preparatory weeks for programming before starting the first semester in CS.
\end{itemize}

\section{BBI recommendations for the near future}\label{sec:recommendations}

\begin{figure}[h]
\centering
\begin{tikzpicture}
\node[draw,circle,minimum size=22mm] (S) at (0,0) {School};
\node[draw,circle,minimum size=22mm] (St) at (5,0) {CS Studies};
\node[draw,circle,minimum size=22mm] (A) at (10,0) {CS Academia};
  
\draw[->, line width=0.7mm] (S) to (St);
\draw[->, line width=0.7mm] (St) to (A);

\node[draw,text width=1.5cm,align=center,dashed] (S) at (2.4,1.5) {Balanced Flow};
\node[draw,text width=2.2cm,align=center,dashed] (S) at (5,2.5) {Balanced\\ Environment};
\node[draw,text width=1.5cm,align=center,dashed] (S) at (7.6,1.5) {Balanced Flow};
\end{tikzpicture}
\caption{We propose action items for (i) a gender-balanced flow of students from the school to CS studies, (ii) a gender-balanced student environment in our department,
and (iii) a gender-balanced flow of graduates from the CS studies to the CS Phd program, and eventually CS academia.}
\label{fig14}
\end{figure}
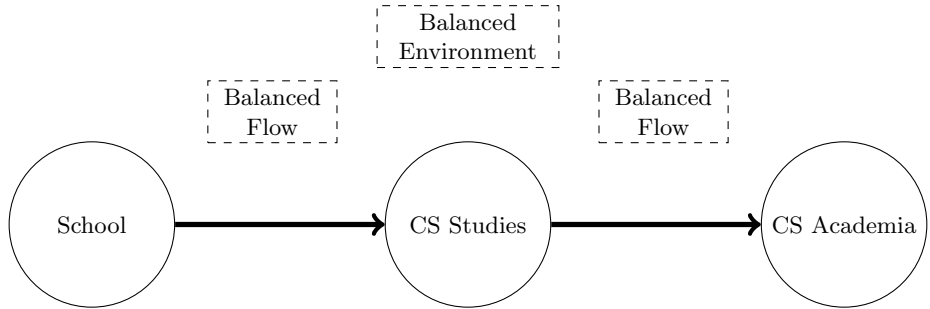

Throughout this report we presented a variety of
 action items for improving gender balance in our CS department. We have motivated these action items using the findings from the discussions with the participants.
We now summarize those actions that BBI recommends for the near future. These actions aim to achieve a balanced flow
of students into CS studies, a balanced environment within the CS department, and a balanced flow towards CS academia.
Figure \ref{fig14} gives a schematic representation of these three categories of actions, which are listed below.

\begin{itemize}
\item[] \textbf{Action items for a balanced flow of students into CS.}
\begin{itemize}
\item[--] Student-ambassadors of all genders to present CS to high school and middle high school students.
\item[-] Highlight problem solving aspect of CS, flexible and interesting program, job opportunities. CS is something that everyone can do.
\end{itemize}
\end{itemize}
\begin{itemize}
\item[] \textbf{Action items for a balanced CS study environment.}
\begin{itemize}
\item[--] Organize social events where senior students offer tips and share experiences and success stories with junior students. 
\item[--] Have mandatory videos with examples of unaccepted behaviors (e.g, inappropriate jokes, stalking, etc).
\item[--] Advise faculty members to ensure that lectures and assignments accommodate a variety of interests.
\item[--] Advise faculty members to ensure that colloquiums are not transformed into boys-clubs.
\item[--] Increase the number of female TAs.
\item[--] Explore the opportunity to have a few preparatory weeks for programming before starting the first semester in CS.
\end{itemize}
\end{itemize}
\begin{itemize}
\item[] \textbf{Action items for a balanced flow of candidates into CS academia.}
\begin{itemize}
\item[--] Hold events that clearly describe the academic career and the PhD program in CS.
\item[--] Advise faculty members to move PhD applications earlier at the academic year.
\end{itemize}
\end{itemize}

\section{Related Work}\label{sec:relatedwork}

The discussion sessions with the students helped us identify action items to achieve (i) a balanced flow
of students into CS studies, (ii) a balanced environment within the CS department, and (iii) a balanced flow towards CS academia (i.e., PhD and beyond). 
This section discusses how prior work tackles these
three aspects separately. For an extensive overview of initiatives for improving gender balance in CS, we refer the reader to Jaccheri \etal~\cite{jaccheri20}.

Our discussion participants highlighted in
Figure \ref{fig13} that we need to ``deal with the issue [of gender balance] at a younger age".
A recent aggregated study~\cite{happe21} collects 22 measures and
strategies for CS educators in secondary education to 
sustain the interest of female students in the CS classes.
Our action items for a balanced flow of students into CS are aligned with the proposed measurements in \cite{happe21} that aim to demolish stereotypes. 
Our observations are also aligned with a study~\cite{Alshahrani18} concluding that: family and friends have high impact to the decision of girls to study CS, courses for CS should be introduced earlier at school, one should highlight the problem-solving aspect of CS and surface female role models.
A successful strategy for increasing the percentage of CS female undergraduate students at CMU (close to $50\%$ was the percentage of female students that entered the CS program in 2018)
is presented in \cite{Frieze19}. 
Two points of this strategy are that the curriculum does not have to change to become more ``female-friendly",
and that it is important to promote cultural changes
within the institution (e.g., create entry level courses for students with no prior programming experience, increase the visibility of women, break stereotypes).
These points address two main reasons~\cite{Varma10}
for the low enrollment of
female students in CS programs:
``bias in early socialization" and ``anxiety towards technology". 
A more recent paper~\cite{zlavi21} refines those reasons into 
three categories: social (e.g., stereotypes), educational (e.g., unattractive CS learning environment),
and labor market (e.g, unattractive jobs). Then the authors
present ways to address those reasons, by communicating 
different perspectives of CS and engaging female students
to various CS experiences.

Understanding the culture within the study environment
of a CS department is a prerequisite for decreasing the 
gender gap. CMU employed student interviews~\cite{Fisher97} to inform its strategy for better gender balance.
Margolis \etal~\cite{Margolis00} investigate how the interest of female
students about their CS studies might decline and eventually lead to drop-out.
Rosenstein \etal~\cite{Rosenstein20} report that, within a sample of 200 students, ``$57\%$ were found to exhibit
frequent feelings of the Impostor Phenomenon with a larger fraction of women ($71\%$) experiencing frequent feelings of the Imposter
Phenomenon than men ($52\%$)".
Miller \etal~\cite{Miller21} focus on students with minoritized identities of sexuality and/or gender (MIoSG) in STEM,
and concludes that these students are enduring 
a ``dude culture" that fosters hypermasculinity and suppresses discourses related to sexual orientations other than heterosexuality.
On the positive side, interviewing STEM students, Rainey \etal~\cite{Rainey19} conclude that active teaching may improve the sense of belonging for underrepresented students.
Finally, Lagesen \cite{Lagesen08} interviews 
Malaysian female students, which form around $50\%$ of the student body, to see how their
perception about CS differs from the western culture.

A more limited number of studies have been devoted to 
fostering a gender-balanced flow of students towards PhD and beyond.
For example, Moreno \etal~\cite{Moreno13} interview doctoral CS students
 on the reasons that led them to apply to a PhD program.
 The authors identified five mean reasons: academic career
goal, professional development, career change, employment
opportunity and personal fulfillment. Personal fulfillment was the most popular reason given.

\section{Conclusion}\label{sec:conclusion}

To understand how the gender balance in our CS department can be improved, we organized discussion sessions among 
CS undergraduate students, who shared their thoughts about: the reasons they picked CS for their studies, 
 their current experience with the CS studies, 
 their intention to pursue an academic career in CS, and 
 ways to make the CS community more diverse and inclusive.
From their answers we identified action items for achieving
a balanced flow of students into CS undergraduate program, 
a study environment that embraces diversity, and 
a balanced flow of students into higher levels of the CS academia.
After the completion of the discussion sessions, the students were able to submit their feedback. 
We were pleased to see that they enjoyed the discussion and thought that the questions we asked were important.
The participants also appreciated our effort to use neutral and not-offensive language for the questions
and the discussion that they triggered. 

\begin{figure}[t]
\centering
\includegraphics[scale=0.07]{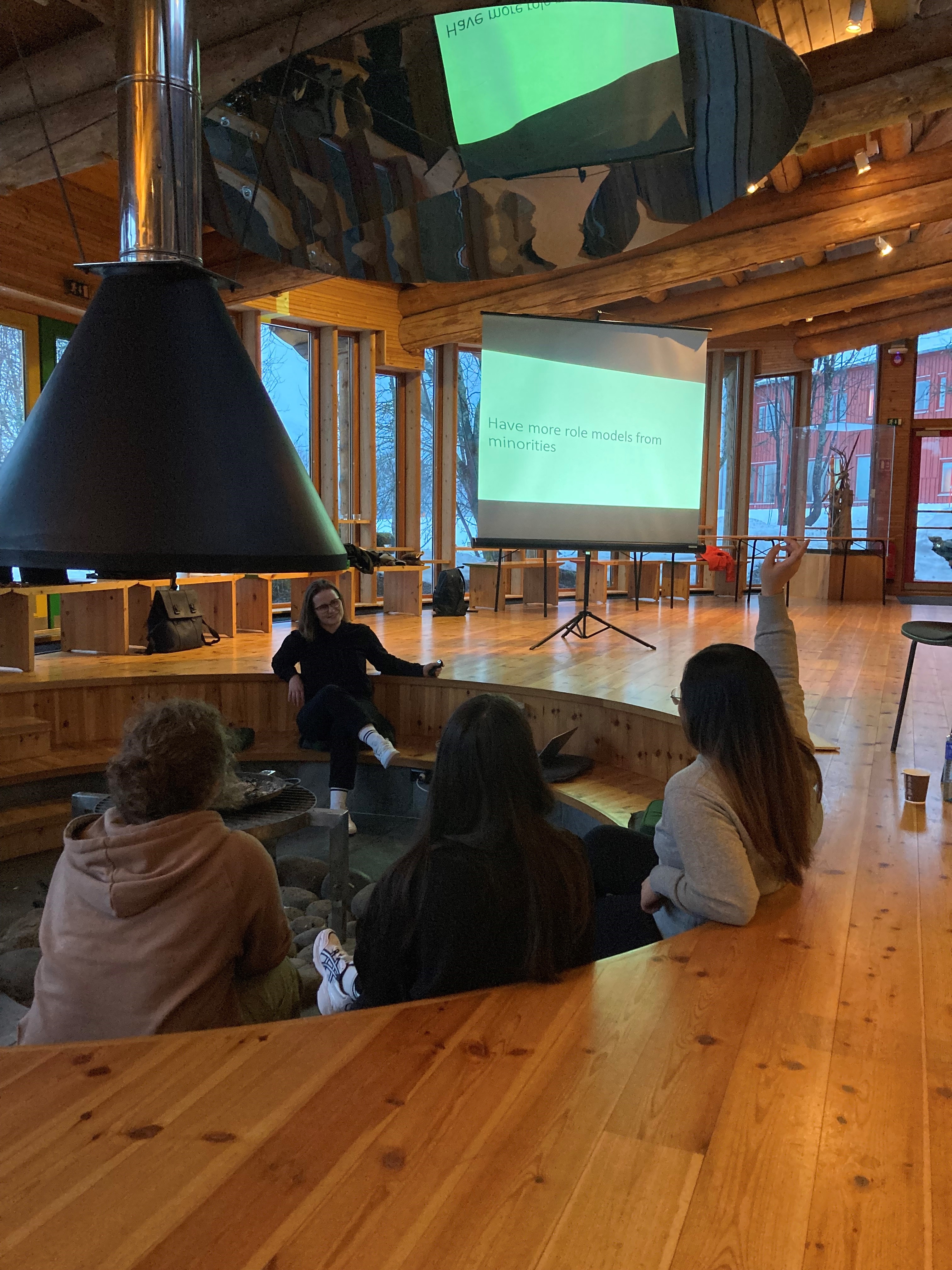}
\caption{A discussion session at Árdna, UiT.}
\label{fig15}
\end{figure}

\section*{Acknowledgements}

We would like to thank Lilli Mittner for recommending 
Árdna for holding the discussion sessions and for giving
inspiration for the discussion questions.
Lynn Nygaard gave us inspiration for these questions, too.
We also thank Melina Duarte for providing network support for BBI within UiT, and Ingeborg Owesen for providing network support for BBI within BalanseHub. Finally, we
are grateful to the administration of the CS department
and the members of BBI for their help in organizing the discussion sessions.
This work has been partially supported by the COST Action CA19122, from the
European Network for Gender Balance in Informatics,
and by
the NFR grant 321075 for BBI.

\bibliographystyle{splncs04}
\bibliography{biblio}

\end{document}